\documentclass[num-refs]{wiley-article}
\usepackage[utf8]{inputenc}
\usepackage{lipsum}
\usepackage{booktabs}
\PassOptionsToPackage{hyphens}{url}
\usepackage{hyperref}
\usepackage{float}
\newcommand*{\restartrowcolors}{%
  \ifhmode\unskip\fi
  \vadjust{%
    \global\rownum=0 %
  }%
}
\usepackage{xcolor}
\usepackage{pgfplots}
\usepackage{subcaption}

\newcommand{\opcode}[1]{\texttt{#1}}
\newcommand{\intrinsic}[1]{\texttt{#1}}
\usepackage[nounderscore]{syntax}
\AtBeginEnvironment{grammar}{\small}

\usepackage{marginnote}
\hypersetup{
    colorlinks,
    linkcolor={red!50!black},
    citecolor={blue!50!black},
    urlcolor={black}
}

\usepackage{todonotes}
\usepackage{microtype}
\usepackage{CJKutf8}
\usepackage[per-mode=symbol]{siunitx}
\usepackage{pgfplots}
\usepackage{tikz}
\usepackage{tikzscale}
\usetikzlibrary{shapes,arrows}

\usepackage{etoolbox}
\usepackage{xcolor}
\usepackage{listings}
\usepackage{tcolorbox}
\lstdefinestyle{customcpp}{%
  belowcaptionskip=1\baselineskip,
  breaklines=true,
  xleftmargin=\parindent,
  language=C++,
  showstringspaces=false,
  basicstyle=\linespread{0.4}\footnotesize\ttfamily,
  keywordstyle=\bfseries\color{green!40!black},
  numberstyle=\tiny,
  commentstyle=\itshape\color{purple!40!black},
  identifierstyle=\bfseries\color{black},
  stringstyle=\color{red},
  emph={int,char,double,float,unsigned},
  emphstyle=\color{blue},
  morekeywords={uint64_t,uint32_t,__m256i,__m128i,UINT64_C},
}

\lstdefinelanguage{JavaScript}{
  keywords={typeof, new, true, false, catch, function, return, null, catch, switch, var, if, in, while, do, else, case, break},
  keywordstyle=\color{blue}\bfseries,
  ndkeywords={class, export, boolean, throw, implements, import, this},
  ndkeywordstyle=\color{darkgray}\bfseries,
  identifierstyle=\color{black},
  sensitive=false,
  numberstyle=\color{red}\ttfamily,
  comment=[l]{//},
  morecomment=[s]{/*}{*/},
  commentstyle=\color{purple}\ttfamily,
  stringstyle=\color{blue}\ttfamily,
  morestring=[b]',
  morestring=[b]"
}

\lstdefinestyle{customjavascript}{%
   language=JavaScript,
   backgroundcolor=\color{white},
   extendedchars=true,
   basicstyle=\linespread{0.4}\footnotesize\ttfamily,
   showstringspaces=false,
   showspaces=false,
   numberstyle=\footnotesize,
   tabsize=2,
   breaklines=true,
   showtabs=false}
\definecolor{bblue}{HTML}{4F81BD}
\definecolor{rred}{HTML}{C0504D}
\definecolor{ggreen}{HTML}{9BBB59}
\definecolor{ppurple}{HTML}{9F4C7C}

\newcommand{\asciicharacter}[1]{`\texttt{#1}'}

\title{Parsing Millions of DNS Records per Second}
\author[1]{Jeroen Koekkoek}

\author[2]{Daniel Lemire}

\affil[1]{NLnet Labs, Science Park 400, 1098 XH Amsterdam, The Netherlands}
\affil[2]{Data Science Research Center, Universit\'e du Qu\'ebec (TELUQ), Montreal, Quebec, H2S 3L5, Canada}

\corraddress{Jeroen Koekkoek, Oldenzaal, The Netherlands}
\corremail{jeroen@koekkoek.nl}

\corraddress{Daniel Lemire, Data Science Research Center, Universit\'e du Qu\'ebec (TELUQ), Montreal, Quebec, H2S 3L5, Canada}
\corremail{daniel.lemire@teluq.ca}

\fundinginfo{Natural Sciences and Engineering Research Council of Canada, Grant Number: RGPIN-2024-03787}

\runningauthor{Jeroen Koekkoek and Daniel Lemire}
\begin{document}
\maketitle

\begin{abstract}
The Domain Name System (DNS) plays a critical role in the functioning of the Internet. It provides a hierarchical name space for locating resources. Data is typically stored in plain text files, possibly spanning gigabytes. Frequent parsing of these files to refresh the data is computationally expensive: processing a zone file can take minutes.

We propose a novel approach called simdzone to enhance DNS parsing throughput. We use data parallelism, specifically the Single Instruction Multiple Data (SIMD) instructions available on commodity processors.  We show that we can multiply the parsing speed compared to state-of-the-art parsers found in Knot DNS and the NLnet Labs Name Server Daemon (NSD). The resulting software library replaced the parser in NSD.
\end{abstract}

\section{Introduction}
The Domain Name System (DNS) was proposed in 1983 by Mockapetris~\cite{rfc882,mockapetris1988development}. It provides a distributed hierarchical name space.  The name space is organized in a tree structure where each node and leaf have an associated label and resource set. 
The root domain (represented by a dot, `.') is at the top.
Beneath the root, there are top-level generic domains (TLDs) like `.com', `.org', and country-code TLDs (e.g., `.ca' for Canada). The unique combination of labels from the leaf to the root forms a domain name (e.g., \texttt{www.example.com.}) that names resources.

Resource records (RRs), simply referred to as record hereafter, are the fundamental units of information in the DNS. A record contains the name, type, class, time to live (TTL), length of the resource data and the resource data (RDATA) itself.

The combination of type and class defines the meaning of the RDATA. Well-known types include the A and AAAA types, which are used to map domain names to one or more IP addresses. Records can be presented in text using the \emph{presentation format}~\cite{rfc9499}. For example, a record for the domain `\texttt{www.example.com.}' might map it an IPv4 address:
\begin{verbatim}
www.example.com.  3600  IN  A  192.0.2.1
\end{verbatim}
We interpret this line as follows:
 \begin{itemize}
\item Name: \texttt{www.example.com.},
\item Type: A (Address),
\item Class: IN (Internet),
\item TTL: 3600 (1 hour),
\item RDATA: 192.0.2.1.
\end{itemize}
In this entry, \texttt{www.example.com.} is the domain name, \texttt{3600} is the time to live (TTL) in seconds, \texttt{IN} stands for Internet, and \texttt{A} indicates that this is an address record containing an IPv4 address \texttt{192.0.2.1}.
Not all records refer to an IP address as the following example illustrates:
\begin{verbatim}
example.com.  1800  IN  TXT  "v=spf1 ip4:192.0.2.1 -all"
\end{verbatim}
We interpret this line as follows:
 \begin{itemize}
\item Name: \texttt{example.com.},
\item Type: TXT (Descriptive text),
\item Class: IN (Internet),
\item TTL: 1800 (0.5 hour),
\item RDATA: "v=spf1 ip4:192.0.2.1 -all".
\end{itemize}

The domain name space is partitioned and the responsibility for the content is delegated due to the size and frequency of updates. A subtree for which authority is delegated is called a \emph{zone} (usually per organization, etc).
E.g., `\texttt{example.com.}' might correspond to a zone managed by a private company. DNS messages are encoded using the binary format defined in RFC~1035~\cite[section~4]{rfc1034} (henceforth the wire format) when transmitted over the network. However, authoritative DNS data is typically stored on disk in plain text in master files, more commonly referred to as zone files. Zone files and the corresponding text format (henceforth the presentation format) were first introduced in~1987 by RFC~1034~\cite[section~3.6.1]{rfc1034} and RFC~1035~\cite[section~5]{rfc1035}, but their specification has been extended over time.

Zone files can become large (e.g., tens of gigabytes), and parsing these files can become a performance bottleneck. 
To accelerate processing, we can use our processors more efficiently. Most
commodity processors (Intel, AMD, ARM, POWER) support
Single Instruction Multiple Data (SIMD) instructions: these  instructions operate on several words at once, using wide registers. 
For example, most Intel and AMD processors have 256-bit vector registers and they can compare two strings of 32~characters in a single instruction. 
    Recent work on JavaScript Object Notation (JSON) parsing~\cite{rfc8259,keiser2023demand,langdale2019parsing}, notably in the simdjson library,
demonstrates that applying SIMD instructions for parsing structured text can significantly boost performance. JSON has a relatively simple grammar compared to zone files, but there is some conceptual overlap. Can the ideas that made JSON parsing faster be applied to parse zone files faster? We sought to find out by building simdzone, whose name is a play on simdjson,  to achieve a similar performance boost for parsing zone data. We find that the in-memory parsing speed of simdzone is nearly a gigabyte per second---at least three times faster than a competitive solution (Knot DNS).

The SIMD instructions improve the speed at two levels in  our parser: they allow us to quickly index the content, and they accelerate the parsing of specific components in the processing (e.g., time stamps and Pv4 addresses). To our knowledge, simdzone is the first parser to bring SIMD-based parsing techniques to zone files.

\section{Related Work}

To our knowledge, there is no related work on the production of high-performance zone parsers. Much of the academic research regarding zone files relates to security issues. For example, Chandramouli and Rose~\cite{chandramouli2005integrity}
proposed a validation scheme for zone files. Korczy{\'n}ski et al.~\cite{korczynski2016zone} discuss insecure zone-file updates 
  which they refer to as \emph{zone poisoning}. 
Kakarla et al.~\cite{kakarla2022scale} present SCALE, a method for finding standard compliance bugs in DNS nameservers. They developed a tool called FERRET, which uncovered 30~new unique bugs across various implementations.
They examined eight implementations: BIND, NLnet Labs Name Server Daemon (NSD), PowerDNS, Knot DNS, COREDNS, YADIFA, MARADNS, TRUSTDNS. Among the faults they found were server crashes (BIND, COREDNS and TRUSTDNS). %No implementation was fully standard compliant.

Though some DNS system vendors conduct extensive benchmarks~\cite{knothbenchmark}, there is relatively little work in the formal literature. One exception is Lencse~\cite{lencse2020benchmarking} who presents a performance analysis of four authoritative DNS server implementations: BIND, NSD, Knot DNS, and YADIFA. The benchmark is focused on serving the data rather than parsing~\cite{rfc8219}. They find that NSD and Knot DNS can achieve an order of magnitude higher performance than BIND and YADIFA.

There is  related work regarding the high-performance parsing of Internet formats. E.g., XML parsing has received much attention~\cite{van2004constructing,Kostoulas:2006:XSI:1135777.1135796,Cameron:2008:HPX:1463788.1463811},  URL parsing can be accelerated~\cite{nizipli2024parsing}, and we can encode and decode base64 data at high speeds~\cite{mula2018faster,mula2020base64}.

There is also  work regarding JSON parsing.: i.e., Langdale and Lemire presented their simdjson library which can parse gigabytes of JSON per second using branchless routines and vectorization~\cite{langdale2019parsing}.  On commodity processors, the simdjson parser was the first standard-compliant JSON parser to process gigabytes of data per second on a single core.
Our work builds on simdjson~\cite{langdale2019parsing}, in particular with respect to its indexing stage. 
The JSON syntax is practically a small subset of JavaScript which preserves the arrays, objects, numbers, strings, Booleans and null values: e.g., \texttt{\{"name":"Jack","numbers":[11,22]\}}).
%: it keeps four primitive types  (string, number, Boolean, null) and two composed types (arrays and objects). An object is a list of key-value pairs between braces while an array is a comma-separated values The JSON specification has six  \emph{structural characters} (`\texttt{[}', `\texttt{\{}', `\texttt{]}', `\texttt{\}}', `\texttt{:}', `\texttt{,}') delimiting the  structure of objects and arrays. 
The simdjson parser  identifies  the starting location of all JSON nodes (e.g., numbers, strings, null, true, false, arrays, objects)
and all JSON structural characters (`\texttt{[}', `\texttt{\{}', `\texttt{]}', `\texttt{\}}', `\texttt{:}', `\texttt{,}'). 
The parser stores these locations as an array of integer indexes. Given the JSON document \texttt{\{"abc":2000\}}, we might have the indexes~0, 1, 6, 7, 11. A difficulty is to  distinguish the characters that are between quotes: it is solved without branching. The simdjson  leverages SIMD instructions to process large blocks (64~bytes) of characters. These blocks are transformed in 64-bit words acting as bitsets from which indexes are extracted.
Though the simdjson approach is designed for commodity processors, there are alternative methods on specialized processors~\cite{10.1145/3533737.3535094,hahn2022raw,peltenburg2021tens,10.14778/3377369.3377372}. For simdzone, we focus  on commodity processors.

In simdjson, the indexing phase covers an entire JSON document. 
One significant conceptual difference between simdjson and simdzone is that simdzone indexes smaller blocks---and not the whole zone file.

\section{Zone Files}

Zones are typically stored in text files containing a sequence of records in the presentation format: a concise tabular serialization format with provisions for more convenient editing by hand. While name servers implement slightly different dialects due to the ambiguity of the specification, the presentation format provides sufficient interoperability in practice.

Zone files consist of a sequence of entries made up of a list of white space separated items. An entry is terminated by a newline, but parentheses can be used to continue a list of items across a line boundary. Any line can end with a comment.  Comments start with a semicolon (`\texttt{;}').
See Fig.~\ref{fig:zone-file-example}. A detailed grammar for expressing items is provided in Fig.~\ref{fig:abnf}.

The following entries are defined by RFC~1035~\cite{rfc1035} and RFC~2308~\cite{rfc2308}:
\begin{itemize}
\item[] \begin{verbatim}
<blank>[<comment>]
\end{verbatim}
\item[] \begin{verbatim}
$ORIGIN <domain-name> [<comment>]
\end{verbatim}
\item[] \begin{verbatim}
$INCLUDE <file-name> [<domain-name>] [<comment>]
\end{verbatim}
\item[] \begin{verbatim}
$TTL <TTL> [<comment>]
\end{verbatim}
\item[] \begin{verbatim}
<domain-name><rr> [<comment>]
\end{verbatim}
\item[] \begin{verbatim}
<blank><rr> [<comment>]
\end{verbatim}
\end{itemize}
Blank lines, i.e., empty lines or lines consisting only of white space characters and/or a comment are ignored. Control entries (i.e.,  \texttt{\$ORIGIN}, \texttt{\$INCLUDE}, \texttt{\$TTL}) are intended for more convenient editing by hand.
\begin{itemize}
\item \texttt{\$ORIGIN} resets the current origin for relative domain names (domain names without a trailing dot). The directive allows for concise notation of domain names.
\begin{verbatim}
$ORIGIN example.com.
www A 192.0.2.1 ; becomes www.example.com. A 192.0.2.1    
\end{verbatim}
\item \texttt{\$INCLUDE} inserts the specified file in the current location. An optional domain name can be provided to define the origin for relative domain names in the file. The directive allows for templating. E.g. records common across zones can be defined in a single file using relative names and included in other files for consistency.
\item \texttt{\$TTL} resets the TTL for records that omit it.
\end{itemize}
Any entry starting with a dollar sign (`\texttt{\$}') is best considered a control entry to allow for vendor specific control entries.

The remaining entry types are used to define records. The last provided owner (record domain name) is used if an entry omits it and starts with one or more white space characters.
%%  Labels in domain names are expressed as character strings

The \texttt{<rr>} section takes one of the following forms:
\begin{verbatim}
    [<TTL>] [<class>] <type> <RDATA>
    [<class>] [<TTL>] <type> <RDATA>
\end{verbatim}
TTL is a 32-bit decimal integer and mnemonics are used to express class and type. TTL and class default to the last provided values if omitted.

The \texttt{<RDATA>} section depends on the type and class.
A complete presentation of the RDATA section for all record types is beyond our scope, but consider the following examples:
\begin{itemize}
\item The RDATA section for an A (Address) record consists of a single IPv4 address. Typically expressed as four decimal numbers separated by dots without interior spaces.
\begin{verbatim}
www.example.com. A 192.0.2.1
\end{verbatim}
\item The RDATA section for an NS (Name Server) record consists of a single domain name.
\begin{verbatim}
example.com. NS ns1.example.com.
\end{verbatim}
\item The RDATA section for a SOA (Start of Authority) record consists of the domain name for a name server that is the primary source of the zone, a domain name specifying a mailbox, a version number as a 32-bit decimal integer, three 32-bit time intervals in seconds, followed by the minimum TTL field for this zone.
\begin{verbatim}
example.com. SOA ns1.example.com. hm.example.com. (
  2024071301 ; YYYYMMDDNN
             ; Increased by 1 for changes
  3600       ; Refresh (how often to check for updates)
  600        ; Retry (how often to retry after a failure)
  86400      ; Expire (how long to use zone data)
  3600 )     ; Minimum TTL (how long to cache data)
\end{verbatim}
\end{itemize}

Labels in domain names and text literals (e.g., TXT records) are expressed as character strings and allow for use of escape sequences to include characters that otherwise have structural significance. E.g., to include a dot (`\texttt{.}') in the label of a domain name it can be escaped (e.g., `\texttt{\textbackslash{}.}'). If the octet is not a printable character a backslash followed by a decimal number describing the octet is allowed too (e.g., `\texttt{\textbackslash{}008}' for backspace). Values for other field types are expressed using the typical representation, no grammar is provided.

A sequence of bytes may have a different meaning based on location. E.g. if an entry starts with IN it is interpreted as a relative domain name whereas if it is preceded by one or more white space characters, it is interpreted as the mnemonic IN, which denotes the class. If the string \texttt{IN} is quoted, it is a character string and is never interpreted as a mnemonic.

RFC~9460~\cite{rfc9460} introduces a key-value concept for the RDATA section of Service Binding (SVCB) record type and slightly updates item grammar. Parameters may be specified in any order as either \texttt{key=value} or \texttt{key="value"}. The key dictates the type of value along with any particular escaping rules. 
For example, the \texttt{port} parameter takes a port number while the \texttt{ipv4hint} parameter takes an IPv4 address. Neither  \texttt{port} nor \texttt{ipv4hint}  allow character escaping. In contrast, the \texttt{alpn} parameter takes a comma-separated list of character strings and hence allows for escaping. Consider the following example:
\begin{verbatim}
example.com.   SVCB   16 foo.example.org. (
                         alpn=h2,h3-19 mandatory=ipv4hint,alpn
                         ipv4hint=192.0.2.1
                         )
\end{verbatim}

Name servers might lack support for some record types. To improve interoperability, RFC~3597~\cite{rfc3597} introduces a generic notation. Instead of using a mnemonic, the type can be specified using the word \texttt{TYPE} directly followed by the type code. E.g. \texttt{TYPE1} and \texttt{A} are equivalent. Similarly, the class can be specified using the word \texttt{CLASS} directly followed by the class code. E.g. \texttt{CLASS1} and \texttt{IN} are equivalent as well. The RDATA section in generic notation is specified using the special token \texttt{\textbackslash{}\#} followed by a 16-bit decimal integer for the length of the RDATA and the RDATA in hexadecimal encoding. For example, to map \texttt{www.example.com.} to \texttt{192.0.2.1}:
\begin{verbatim}
    www.example.com. 3600 CLASS1 TYPE1 \# 4 C0000201
\end{verbatim}
% Jeroen: I would like this figure to directly follow the above text if possible.
% Daniel: You can replace '\begin{figure}' by '\begin{figure}[H] in the next line... but this is not recommended. LaTeX will then force the figure to appear at the next available spot, possibly leaving a lot of white space. LaTeX relies on white space optimization to optimize the layout, if you force figure placement, you break the algorithm and may get bad results. Still, you can do it. Be mindful that a journal will have its own copy editing, page format, fonts and so forth, and they will almost surely not abide by such constraints (and they actively discourage it).
% Jeroen: Blame it on my inexperience with submitting papers :) Let's keep it as is.
\begin{figure}
    \centering
    \small\begin{tcolorbox}
    \begin{verbatim}
$ORIGIN example.com.  ; Defines the origin of the zone

; Free standing @ is replaced by origin
@ 3600 SOA ns1.example.com. hm.example.com. ( 2024071301 ; YYYYMMDDNN
    ; Increased by 1 for changes
    3600 ; Refresh (how often to check for updates)
    600 ; Retry (how often to retry after a failed refresh)
    86400 ; Expire (how long to use zone data)
    3600 ) ; Minimum TTL (how long to cache data)

; Name server records
ns1 86400 NS ns1.example.com.
ns2 86400 NS ns2.example.com.

; Mail server record
mx 10 mail.example.com.

; Web server record
www 3600 A 192.168.1.100        
    \end{verbatim}
    \end{tcolorbox}
    \caption{Example of a simple zone file}
    \label{fig:zone-file-example}
\end{figure}

\begin{figure}
\centering
\begin{tcolorbox}
\begin{minipage}[b]{0.9\columnwidth}
\begin{grammar}
<non-special> ::= OCTET - NUL, WSP, LF, DQUOTE, ";", "(", ")", and "\\"

<non-digit> ::= OCTET - NUL and DIGIT

<dec-octet> ::= ("0" / "1") 2DIGIT / "2" (("0"-"4") DIGIT) / ("5" ("0"-"5"))

<escaped> ::= "\\" (non-digit / dec-octet)

<contiguous> ::= 1*(non-special / escaped)

<non-dquot> ::= OCTET - NUL, DQUOTE, and "\\"

<quoted> ::= DQUOTE *(non-dquot / escaped) DQUOTE

<char-string> ::= contiguous / quoted
\end{grammar}
\end{minipage}
\end{tcolorbox}
\caption{Augmented Backus–Naur form (ABNF) grammar for items in resource records\label{fig:abnf}}
\end{figure}

\section{Data Parallism}
\label{sec:advancedinstructions}

Commodity processors offer advanced Single Instruction Multiple Data (SIMD) instructions (e.g., SSE4.2, AVX2, etc.) to process multiple data elements simultaneously using a single instruction. While optimizing compilers can make use of them, a deliberate design of the algorithms by the programmer remains needed in most cases. In C/C++, SIMD instructions are available through special functions called intrinsics (see Table~\ref{ref:simdinstructions}) 

Most intrinsics are straightforward. For example, the \intrinsic{\_mm256\_cmpeq\_epi8} intrinsic takes two 32-byte registers and outputs a new 32-byte register containing of the result of the comparisons of the inputs. A zero byte indicates that the bytes were different whereas a byte value of 255 indicates that the bytes were equal.

The  \opcode{vpshufb}  instruction provides a vectorized table lookup. It works independently on two lanes of 16~bytes.
Given an input register $v$ and a control mask $m$, it outputs new values $(v_{m_0},v_{m_1},v_{m_2},v_{m_3}, \ldots, v_{m_{15}})$ with the additional convention that only the least significant 4~bits of $m_0, m_1, \ldots$ are considered and that if  the most significant bit of $m_i$ is set, then the result is zero (e.g., $v_{128} \equiv 0$).
It has low latency (usually 1~CPU cycle) and a high throughput~\cite{fog2016instruction}. POWER and ARM processors have instructions similar to \opcode{vpshufb} with slightly different conventions.

For parsing tasks, SIMD instructions can be leveraged to compare multiple bytes against a pattern in parallel rather than sequentially. Traditional parsing might involve many if-else statements or loops to check each byte. Using SIMD, we can reduce the number of branches, which are costly in terms of CPU pipeline flushes.

\begin{table}
\caption{\label{ref:simdinstructions}Intel AVX2 intrinsics and instructions}%
 \centering
\begin{tabular}{lccp{1.6in}}
\toprule
intrinsic & bits & instruction & description  
\\\midrule
\intrinsic{\_mm256\_or\_si256} & 256 & \opcode{vpor} & bitwise OR \\
\intrinsic{\_mm256\_and\_si256} & 256 & \opcode{vpand} & bitwise AND \\
\intrinsic{\_mm256\_cmpeq\_epi8} & 256 & \opcode{vpcmpeqb} & compare 32 pairs of bytes, outputting 0xFF on equality and 0x00 otherwise \\ 
\intrinsic{\_mm256\_movemask\_epi8} & 256 & \opcode{vpmovmskb} & construct a 32-bit integer from the most significant bits of 32~bytes\\
\intrinsic{\_mm256\_shuffle\_epi8} & 128 & \opcode{vpshufb} & shuffle two lanes of 16 bytes \\
\intrinsic{\_mm256\_loadu\_si256} & 256 &  \opcode{vmovdqu}& load 32 bytes from memory into a vector register \\
\intrinsic{\_mm256\_storeu\_si256} &  256 &  \opcode{vmovdqu}& write a 32-byte vector register to memory \\ 
\bottomrule
\end{tabular}
\restartrowcolors{}
\end{table}

\section{Architecture}
\label{sec:archicture}
A complete description of the simdzone parser would be overly technical. Instead, we describe the architecture, and refer the interested reader to our freely available source code.

Like simdjson~\cite{langdale2019parsing}, simdzone works in two stages: an indexing (stage~1) and a parsing (stage~2) stage. Stage one seeks to determine the location of structural characters (blank and special characters as presented in Fig.~\ref{fig:blankspecial}) and the start and end location of each \emph{item} (character-strings and sequences of non-structural characters~\cite{rfc1035}). At a low level, indexing operates on 64-byte blocks and uses vectorized classification to identify structural characters. The relative indexes are stored in a bitmask and binary logic is used to identify items. The indexes of all structural characters, the first character of each item and the delimiting character for each item are retained. See Fig.~\ref{fig:identifystructure}.

\begin{figure}
{
\begin{center}
\begin{minipage}[b]{0.8\textwidth}
\begin{verbatim}
example.com.  3600  IN  TXT  "v=spf1 ip4:192.0.2.1 -all"  ; SPF : input data
_______________________________________________________________1: newline
__________________________________________________________1_____: semicolon
__________________________________________________________11111_: in_comment
_____________________________1_________________________1________: quoted
_____________________________11111111111111111111111111_________: in_quoted
____________11____11__11___11_______1_____________1_____11_1____: blank
111111111111__1111__11__111_____________________________________: contiguous
1_____________1_____1___1____1_________________________________1: fields
____________1_____1___1____1___________________________1________: delimiters
\end{verbatim}
\end{minipage}
\end{center}
}
    \caption{Bit representation of a 64-byte zone file input. After computing the `fields' value,
    the indexer would write~0, 14, 20, 24, 29,~63 to the first index array---values corresponding
    to the location of the~1s in `fields' and~12, 18, 22, 27, 55~to the second index array---values corresponding to the location of the~1s in `delimiters'.
    }
    \label{fig:identifystructure}
\end{figure}

Specifically, stage one proceeds by successively loading 64-byte blocks. The data is loaded in a structure as in Fig.~\ref{fig:sixtyfour} where the input is made of two 32-byte wide registers for the AVX2 instruction set. The 64-bit words act as bitsets, where each bit corresponds to a byte in the input. The \texttt{newline}, \texttt{backslash}, \texttt{quoted} and \texttt{semicolon} words are initialized by comparing the 64-byte input with the corresponding ASCII characters (\asciicharacter{\textbackslash{}n},  \asciicharacter{\textbackslash{}}, \asciicharacter{"} and \asciicharacter{;}) using fast SIMD comparisons with \opcode{vpcmpeqb} followed by a conversion to a 64-bit register with \opcode{vpmovmskb}. That is, as little as two instructions are needed to identify the locations of the newlines, backslashes, etc.

These straightforward comparisons are used for characters we must uniquely identify for bit manipulation. Characters in the same class that are equally important are located using \emph{vectorized classification}. We mark the \asciicharacter{ }, \asciicharacter{\textbackslash{}t} and \asciicharacter{\textbackslash{}r} with the \texttt{blank} word and the parentheses, the null character and newline with the \texttt{special} word. The \asciicharacter{\textbackslash{}r} character is not technically part of the class \texttt{blank}, we treat it as such for correct length calculation on machines where a newline consists of \texttt{\textbackslash{}r\textbackslash{}n} rather than \texttt{\textbackslash{}n}.

Specifically, we use the least significant 4~bits of each byte to lookup a byte value in a table and compare the result with the original byte value: if they match then we know that we have marked the characters. We present the tables in Fig.~\ref{fig:blankspecial}.
Our generic function is presented in
Fig.~\ref{fig:generic} for the AVX2~case: 
we identify blank characters
by calling \texttt{simd\_find\_any\_8x64(input, blank)} and special characters by calling 
\texttt{simd\_find\_any\_8x64(input, special)}. These functions process 
64~bytes of input and they return a 64-bit word where the bits
are set to identifying the target
characters (blank or special).

The rest of the processing follows by doing scalar operations over the words. E.g., we need to compute the bitwise AND NOT with the word indicating the presence of quoted or commented characters. Determining the location of quoted content is branch-free, but comments complicate indexing. If a semicolon appears, a scalar loop is used to correctly discard semicolons within quoted sections and quotes within comments.

\begin{figure}
    \centering
    \begin{tabular}{c}
\begin{lstlisting}[style=customcpp]
struct block {
  simd_8x64_t input;
  uint64_t newline;
  uint64_t backslash;
  uint64_t escaped;
  uint64_t comment;
  uint64_t quoted;
  uint64_t semicolon;
  uint64_t in_quoted;
  uint64_t in_comment;
  uint64_t contiguous;
  uint64_t follows_contiguous;
  uint64_t blank;
  uint64_t special;
};
\end{lstlisting}
    \end{tabular}
    \caption{Structure in C containing a 64-byte block}
    \label{fig:sixtyfour}
\end{figure}

\begin{figure}
    \centering
    \begin{tabular}{cc}
\begin{lstlisting}[style=customcpp]
simd_table_t blank = SIMD_TABLE(
  0x20, // 0x00 :  " "
  0x00, // 0x01
  0x00, // 0x02
  0x00, // 0x03
  0x00, // 0x04
  0x00, // 0x05
  0x00, // 0x06
  0x00, // 0x07
  0x00, // 0x08
  0x09, // 0x09 : "\t"
  0x00, // 0x0a
  0x00, // 0x0b
  0x00, // 0x0c
  0x0d, // 0x0d : "\r" 
  0x00, // 0x0e
  0x00   // 0x0f
);
\end{lstlisting} &
\begin{lstlisting}[style=customcpp]
simd_table_t special = SIMD_TABLE(
  0x00, // 0x00 : "\0"
  0x00, // 0x01
  0x00, // 0x02
  0x00, // 0x03
  0x00, // 0x04
  0x00, // 0x05
  0x00, // 0x06
  0x00, // 0x07
  0x28, // 0x08 :  "("
  0x29, // 0x09 :  ")"
  0x0a, // 0x0a : "\n"
  0x00, // 0x0b
  0x00, // 0x0c
  0x00, // 0x0d
  0x00, // 0x0e
  0x00   // 0x0f
);
\end{lstlisting} 
    \end{tabular}
    \caption{Tables used for vectorized classification of the blank and special characters}
    \label{fig:blankspecial}
\end{figure}

\begin{figure}
    \centering
    \begin{tabular}{c}
\begin{lstlisting}[style=customcpp]
typedef struct { __m256i chunks[2]; } simd_8x64_t;

typedef uint8_t simd_table_t[32];

uint64_t simd_find_any_8x64(
  simd_8x64_t *simd, simd_table_t table)
{
  __m256i t = _mm256_loadu_si256((const __m256i *)table);

  __m256i r0 = _mm256_cmpeq_epi8(
    _mm256_shuffle_epi8(t, simd->chunks[0]), simd->chunks[0]);
  __m256i r1 = _mm256_cmpeq_epi8(
    _mm256_shuffle_epi8(t, simd->chunks[1]), simd->chunks[1]);

  uint64_t m0 = (uint32_t)_mm256_movemask_epi8(r0);
  uint64_t m1 = (uint32_t)_mm256_movemask_epi8(r1);

  return m0 | (m1 << 32);
}
\end{lstlisting}
    \end{tabular}
    \caption{Generic C function to identify target characters in a block of 64~characters using vectorized classification with AVX2 intrinsics}
    \label{fig:generic}
\end{figure}

When loading blocks of data (i.e., 64~bytes), it is convenient not to have to check for end-of-stream loads---where fewer bytes may be available. Thus, simdzone requires input buffers to be sufficiently large to ensure optimized operations can safely load blocks of input data without reading past the buffer limit. This requirement is of no concern to the user of simdzone when parsing files as simdzone manages input buffers.

The second stage (parsing) processes all items and structural characters. The parsing stage follows the indexing stage, but the two stages are interleaved. Interleaving is required because zone files can be large. However, records are relatively small and are parsed independently. Small windows that often fit in CPU cache are indexed and parsed successively. As a consequence, we must handle with care partially indexed tokens: we use an adaptive window that grows as needed up to a pre-determined maximum ($\approx$\SI{1}{\mega\byte}).

A zone file may contain the \$INCLUDE control directive. 
If such a control directive is encountered, the parser is expected to parse the data in the specified file before parsing the remaining contents in the current file. Indexer state is therefore managed on a per file basis.

Indexes are written to a \emph{tape}, a fixed size, pre-allocated region to avoid allocating memory for each item. While a single tape suffices to implement the logic, delimiting indexes are written to a secondary tape. Deciding if a delimiting index must be discarded introduces extra branches and data dependencies in the hot path. E.g., indexes for delimiting white space characters must be discarded whereas indexes for newline characters have structural significance and must be retained. Writing delimiting indexes to a separate tape allows for discarding the delimiting index unconditionally. During the second stage the item length in bytes is determined by subracting the starting index from the delimiting index. Knowing the length beforehand simplifies parsing the many different data types tha can appear in zone files. E.g., it is not required to account for delimiters in the input when parsing either quoted or contiguous character strings.

%
% Jeroen: Should we mention that records have a maximum size of 65535 bytes
%         (anything bigger results in an error) and that we use that for pre-allocated
%         output buffers?
% Daniel: If the specification says that records are limited to 65535 bytes, it might be interesting to specify it. If it is a limit that simdzone adopted, it seems like an implementation detail that is not super exciting. It could be omitted.
% Jeroen: RRs are always limited to 65535 bytes. The rdlength is an unsigned 16-bit integer
%         and this limit is
%
Given the record type, the data has a predefined layout, though records may contain any type of data as long as it does not exceed 65535 bytes in size. This maximum is imposed by the wire format where a 16-bit unsigned integer is used to specify the data length. We use this hard limit to preallocate the output buffer and avoid memory allocations and simplify buffer management.

\subsection{Domain Names}
There are many data types.
We optimized the processing of the most common ones.
In particular,
domain names account for much of the data, especially since frequently used records reference other domain names. E.g. RDATA for NS, SOA, MX, CNAME and SVCB records all contain one or more domain names. Much like the indexing stage, parsing of domain names is typically implemented as a scalar loop operating on individual bytes, handling escape sequences, label separators (dots) and label requirements. 
We must parse the text representation into the wire format~\cite{rfc1035}:
\emph{each label is represented as a one octet length field followed by that number of octets}.
If we compare the wire format of domain names to the presentation format, the main difference between the two is that the dots in the presentation format are replaced by the length of the label. E.g. \texttt{www.example.com.} is encoded in wire format as \texttt{3www7example3com0} (where the digits represent the actual number, not the ASCII equivalent). Most of the data can therefore be copied without modification. While escape sequences may occur, domain names rarely include characters not part of the preferred name syntax. See Fig.~\ref{fig:preferrednamesyntax}. Registrants typically select domain names that are easy to remember. Consequently, domain names often exceed 16~bytes, but not 32~bytes. We apply the logic used to optimize the indexing stage  to the processing of domain names. We load and store 32~bytes of data unconditionally. We use fast SIMD comparisons to check for \asciicharacter{.} and \asciicharacter{\textbackslash{}} in the input and the results are converted to bitmasks. If escape sequences are detected, i.e., the bitmask for \asciicharacter{\textbackslash{}} is non-zero, the parser unescapes one sequence and starts just after the escape sequence the next iteration. The parser then proceeds to copy and shift the dots bitmask by one to detect null-labels followed by a scalar loop over the indexes to replace the dots in the output by the length of the label.

\begin{figure}
\centering
\begin{tcolorbox}
\begin{minipage}[b]{0.9\columnwidth}
\begin{grammar}
<domain> ::= <subdomain> | " "

<subdomain> ::= <label> | <subdomain> "." <label>

<label> ::= <letter> [ [ <ldh-str> ] <let-dig> ]

<ldh-str> ::= <let-dig-hyp> | <let-dig-hyp> <ldh-str>

<let-dig-hyp> ::= <let-dig> | "-"

<let-dig> ::= <letter> | <digit>

<letter> ::= any one of the 52 alphabetic characters A through Z in
upper case and a through z in lower case

<digit> ::= any one of the ten digits~0 through~9
\end{grammar}
\end{minipage}\end{tcolorbox}
\caption{Preferred name syntax from RFC~1035 section~2.3.1
\label{fig:preferrednamesyntax}}
\end{figure}

\subsection{Record Types}
Records must contain an identifiable TYPEs (RTYPE) followed by a data payload (RDATA). Typically, given the type, the RDATA layout is predictable. This knowledge can be leveraged to expect the right token and fallback to a slower path in uncommon cases. The optimized (common) path can be inlined while the slower path resides in a function. Binary size is hereby reduced as much as possible while token extraction is as fast as can be. This knowledge also allows for calling the correct parser functions in order, eliminating the need for calling more generalized parser functions based on a descriptor table.
There is a finite number of RTYPE values: e.g., ``A'', ``AAAA'', ``AFSDB'', ``APL'', ``CAA'', ``CDS'', ``CDNSKEY'', ``CERT'', ``CH'', ``CNAME'', ``CS'', ``CSYNC'', ``DHC'', etc. We need to support 67~types. Identifying the type quickly, with minimal branching is important. For this purpose, we  designed a perfect hash function~\cite{10.5555/331120.331202,czech1997perfect}. We use a 256-byte table containing either an empty value. Given the input string, we load it into two 8-byte registers---including some possible trailing content. We convert it to ASCII upper case with a binary operation. We set to zero all bytes after the length of the string, by loading a mask from a precomputed table and computing a bitwise AND.
We then apply a hash function on the first eight bytes (see Fig.~\ref{fig:hash}) and take the 8-bit value as an index in the 256-element table. The hash function uniquely distinguishes 
between the existing types: no two type has the same 8-bit value. If the recovered element is empty, we know that the type is unrecognized. Otherwise, we check that we have the correct type by an efficient comparison and we use the resulting code to branch into the corresponding processing.
Though our approach assumes that the types are fixed, we can update the code when new types are added with scripts.

\begin{figure}
    \centering
    \begin{tabular}{c}
\begin{lstlisting}[style=customcpp]
uint8_t hash(uint64_t prefix) {
  uint32_t value = (uint32_t)((prefix >> 32) ^ prefix);
  return (uint8_t)((value * 3523264710ull) >> 32);
}
\end{lstlisting} 
    \end{tabular}
    \caption{Function used as part of our perfect hash function to recognize resource types}
    \label{fig:hash}
\end{figure}

\subsection{Binary Data}
Binary data is presented in text using some of the conventional binary encoding schemes~\cite{rfc5155} such base16 or base64. For example, RDATA in generic notation is presented using base16 (hexadecimal) encoding and the signature in RRSIG records is presented using base64 encoding. Encoding and decoding of base64 data can be done at high speeds~\cite{mula2018faster,mula2020base64} and prior work on base16 decoding demonstrates similar improvements can be achieved there. To speedup loading of DNSSEC signed zones, we integrated an optimized base64 decoder~\cite{aklompbase64}, though only the scalar version. The base64 decoder served as a template for integrating an optimized base16 decoder~\cite{client9stringencoders}. We implemented an optimized base32hex decoder for the SSE4.2 and AVX2 instruction sets to improve parsing of NSEC3 records.

\section{Experiments}

We seek to benchmark the in-memory parsing speed of zone files.
To our knowledge, there has been no benchmark of zone file parsing previously. Previous work~\cite{lencse2020benchmarking} has focused on data queries---based on an already loaded database.
Unfortunately, some
popular DNS systems such as BIND and PowerDNS, do not offer a standalone parser: rather they integrate zone file parsing with the construction of the database.
Thankfully, two of the fastest and most popular DNS systems (Knot DNS and NSD) have functions
that merely parse the data from a file. Like simdzone, they allow us
to pass a function as a parameter which receives the parsed records. By passing a trivial function (e.g., one that merely counts records), we can measure the parsing speed itself.

\subsection{Software}

For  benchmarking, we consider three software packages.
\begin{enumerate}
\item Our own simdzone parser\footnote{\url{https://github.com/NLnetLabs/simdzone}}. We use the code version identified by the commit tag \texttt{aab6386}. It is written in C: it consists in over \num{35000}~lines of code and about \num{1000}~distinct functions. Of this total, our tests alone consists in nearly \num{8000}~lines of code. The software follows the high-level architecture presented in \S~\ref{sec:archicture}. Additionally, some processor-dependent code is organized in three different kernel of functions: fallback, westmere and haswell. These kernels are specific to processors having no supported SIMD instructions, to x64 processors equivalent to Intel~Westmere processors (16-byte SIMD) and to x64 equivalent or superior to Intel~Haswell processors (32-byte SIMD). At runtime, we select the best kernel for the current processor. Further, we have a large set of generic functions which are written in an processor-independent manner, but can be compiled several times, to match different processors.
\item We use Knot DNS version~3.3.4 part of the Ubuntu Linux distribution (version~24.04).
\item We compare against the parser used by NSD up to version~4.9.\footnote{\url{https://github.com/NLnetLabs/nsd}}---prior to its adoption of simdzone. To help reproducibility, we extracted the parser to its own library and we make it freely available.\footnote{\url{https://github.com/k0ekk0ek/zonec}} It relies on Flex and Bison. We use the versions of Flex and Bison available as part of Ubuntu~24.04 for benchmarking purposes.
\end{enumerate}
Though details differ, all three software packages are used in a similar manner:
\begin{itemize}
    \item A parser structure is constructed and initialized. For example, a default TTL is specified (we use 36000).
    \item A \emph{callback} function is assigned. During the main processing, this function is called each time a record is parsed. This function accepts several values: the name, the type, the class, the  TTL and the RDATA of the record.
    \item We must also assign a filename to the parser. The parser handles file inputs.
    \item We  then call a parsing function which will process the records in sequence and call the assigned \emph{callback} function repeatedly---until the zone file has been completly parsed, or a fatal error is encountered.
    \item When the parsing is completed, we call a closing function which closes the files.
\end{itemize}

Our complete benchmarking framework is also freely available.\footnote{\url{https://github.com/lemire/zone_benchmarks}} 
To help reproducibility, we publish as part of benchmark framework a docker file based on the \texttt{ubuntu:24.04}~image, and we run our benchmarks inside a docker container: docker containers running under a Linux system have practically no overhead except for input/output operations~\cite{10495554,keller2023containers}. Our C software is built using the CMake build system. We build the software in release mode using CMake's defaults (\texttt{-O3 -DNDEBUG}). The
compiler is GCC~13.2---the default compiler under Ubuntu~24.04.

Our software first runs each function once (without timing) and then we run it five times, measuring the average and the minimum time. We include a pure file read function along with the three parsing functions. We verify that the difference between the average and the minimum time is less than 1\% in all parsing functions. The non-parsing file reading function has more variation (about 20\%), but it is also significantly faster.

\subsection{Benchmarking Method}

Zone files are used to persist zone data to disk. While zone files can be copied by means of a file transfer protocol, the presentation format is never used over the network. Thus we seek to measure the speed with which we can deserialize zone files.

A methodological issue is that the zone file parsing functions from 
Knot DNS, NSD, and our own simdzone, assume that the data resides on disk. Disk performance varies greatly, from a few megabytes per second to gigabytes per second. Some recent disks support speeds above ten gigabytes per second. Network bandwidth varies in a similar manner.
To avoid benchmarking the specific disk from our benchmarking system, we
rely on the fact that our operating system buffers the zone files to memory when reading it more than times. We can verify that disk access is no longer a bottleneck by including a disk-read benchmark as part of our software.  Specifically, we verify that we can 
read the zone files at speeds greater than \SI{5}{\giga\byte\per\second} which is several times greater than our parsing speed. Alternatively, we could have used a RAM disk.

We deliberately make our results independent from disk performance. Our benchmark is computational: it depends on the performance our the processor. As a side-effect, it makes our measurements more precise: there is little source of variance. In real-world systems, there are other bottlenecks than pure parsing speed---but these considerations are outside of our scope.

\subsection{Data}
% Jeroen: This section still needs a bit of updating to make it fit in better (probably).
% Daniel: I have cut it down because the early part was repetitive. Let us just state what we did. I think we could extend to other zone files, but two is fair enough.

%There are various types of zones, including:
%\begin{itemize}
%\item Root zone \texttt{.} (single dot), which mostly contains glue records for Top-Level Domains (TLD).
%\item Top-Level Domain (TLD) zones, which mostly contain glue records for second-level domains. TLDs include country code Top-Level Domains (ccTLD) such as \texttt{.nl.} for the Netherlands and generic Top-Level Domains (gTLD) such as \texttt{.com.}.
%\item Second-level domain zones, like \texttt{.example.com.} which mostly contains non-glue data, for example addresses (\texttt{A} and \texttt{AAAA}) where the website for the organization is hosted. Zones for second-level domains typically do not receive many updates and are often small by comparison, though this depends on the registrant. A zone for a well known cloud provider may contain more data than the average TLD zone.
%\end{itemize}
%Thus we picked two important ones: 
We picked two important Top-Level Domain (TLD) zone files:
\texttt{.com} and \texttt{.se}. Both are freely available: from the Centralized Zone Data Service (CZDS) and the Swedish Internet Foundation respectively. Table~\ref{tab:datasets} presents the general characteristics of these zone files.

\subsection{Hardware}

We conduct our tests on a Linux server with two Intel Xeon Gold~6338   (x64 Ice~Lake microarchitecture, 2019) processors with \SI{376}{\giga\byte} of DDR4 (3200\,MT/s) memory. These processors have a maximal frequency of \SI{3.49}{\GHz} with a base
frequency of \SI{2}{\GHz}.
They have \SI{48}{\mega\byte} of last-level cache. Though such a system is capable of running multiple threads at once, all of our code is deliberately single threaded. However, the relatively large amount of memory makes it possible to keep the zone files in cache---practically avoiding disk access.
Our processor supports advanced SIMD instructions (AVX-512) but we limit our implementation to AVX2 instructions: AVX2 is available on almost
all currently sold x64 processors unlike AVX-512.

\begin{table}
\caption{Datasets}
\label{tab:datasets}
\centering
\begin{tabular}{lrrcc}
\toprule
zone & bytes & records & comments & date \\
\midrule
\texttt{.com} & \SI{24.7}{\giga\byte} & \num{412174597} &  0 & May 3 2024\\
\texttt{.se} & \SI{1.4}{\giga\byte} & \num{8509650} &  8 & May 6 2024\\
\bottomrule
\end{tabular}
\restartrowcolors{}
\end{table}

\begin{figure}\centering
 \begin{subfigure}[h]{0.49\textwidth}
 \includegraphics[width=0.99\textwidth]{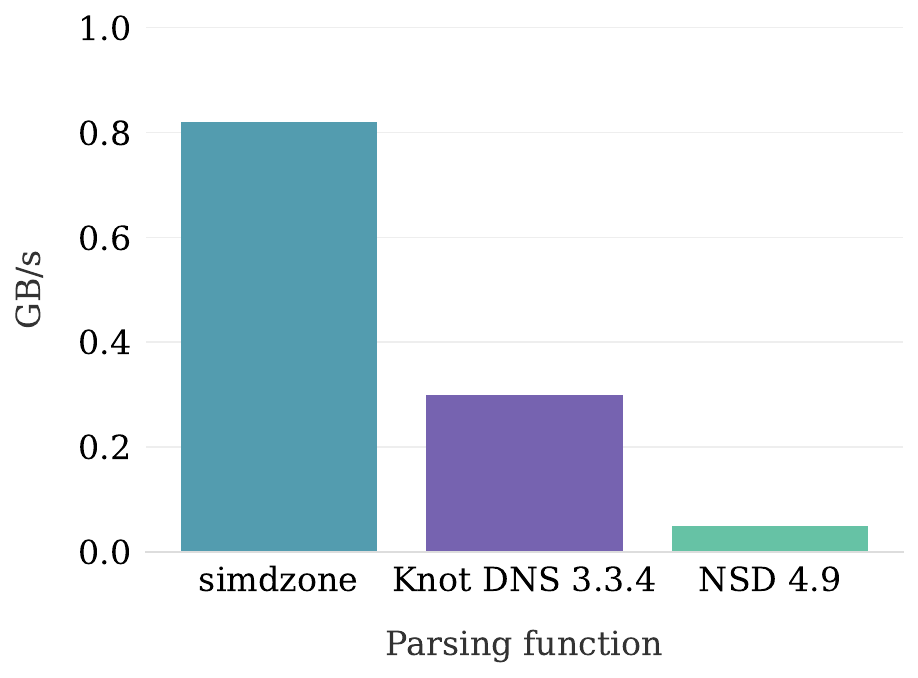}
\caption{\texttt{.com} zone file} \end{subfigure}
 \begin{subfigure}[h]{0.49\textwidth}
 \includegraphics[width=0.99\textwidth]{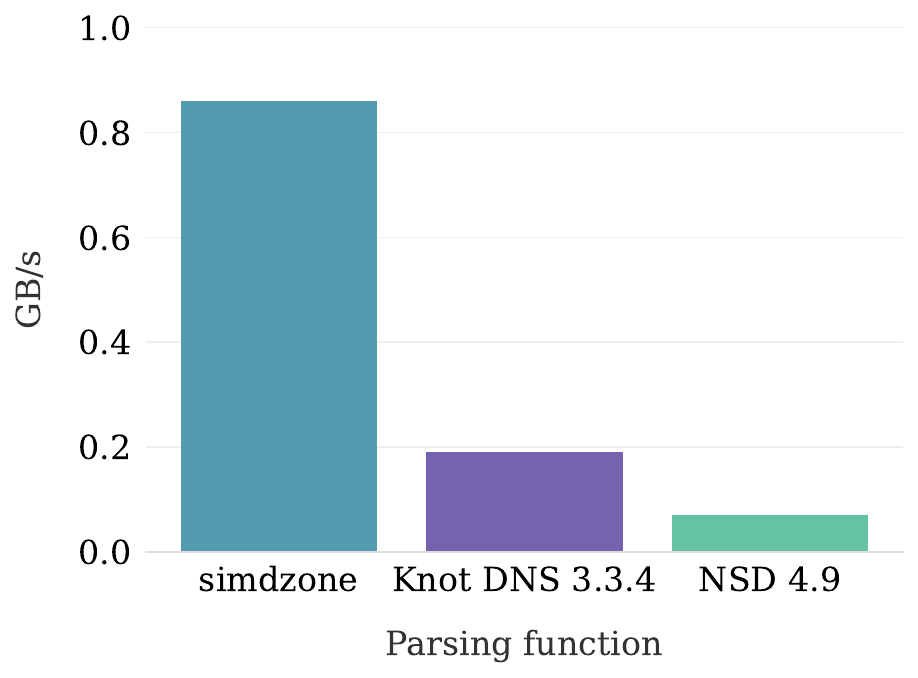}
\caption{\texttt{.se} zone file} \end{subfigure}
\caption{Parsing speed in gigabytes per second\label{fig:speed}}
\end{figure}

\subsection{Results}

Fig.~\ref{fig:speed} presents the parsing speed for the two zone files.
In both instances, the zone parser runs at over \SI{0.8}{\giga\byte\per\second}.
The second fastest parser is Knot~DNS with a performance of \SIrange{0.19}{0.32}{\giga\byte\per\second}: it makes simdzone three to four times faster.
The NSD parser has much lower performance, being about ten times slower
than simdzone: \SIrange{0.05}{0.07}{\giga\byte\per\second}.

To help understand our performance, we use performance counters to record the number of instructions retired by byte as well the number of instructions per cycle and the effective frequency. See Table~\ref{table:counters}. We find that simdzone uses
less than half the number of instructions than Knot DNS, and less than one tenth 
the number of instructions than NSD~4.9. The simdzone parsers also retires
more instructions per cycles than Knot DNS, with a slightly lesser effective frequency.
Thus, overall, the simdzone parser  requires fewer instructions than
a competing parser such as Knot~DNS while being able to retire more instructions
per cycle---at least in these tests. It matches 
our expectation: we require fewer instructions
to process the same data in part because our
software architecture is designed for SIMD instructions. In turn, a single SIMD instruction might
carry the work that would  require  several conventional instructions. Importantly, we reduce significantly the number of needed instructions while maintaining a relatively high rate of instruction execution (instructions per cycle).

\begin{table}\centering
\caption{Performance counters: retired instructions per input byte, instructions retired per cycle, and effective CPU frequency \label{table:counters}}
    \begin{subtable}[h]{0.8\textwidth}\centering
\caption{\texttt{.com}}
\begin{tabular}{lccc}
\toprule
name & instr./byte &  instr./cycle & freq. (\si{\GHz})\\\midrule
simdzone & 12 & 3.5 &  2.8 \\
Knot DNS &  27 & 2.6&  3.1 \\
NSD 4.9 &  204 & 3.3 &  3.49\\\bottomrule
\end{tabular}
\restartrowcolors{}
 \end{subtable}
 
% \restartrowcolors{}
    \begin{subtable}[h]{0.8\textwidth}\centering
\caption{\texttt{.se}}
\begin{tabular}{lccc}
\toprule
name & instr./byte &  instr./cycle & freq. (\si{\GHz}) \\\midrule
simdzone &  11 & 3.4 &   2.8 \\
Knot DNS &  23 & 1.4&  3.2 \\
NSD 4.9 &  127 & 2.9& 3.49 \\\bottomrule
\end{tabular}
 \end{subtable}
% \restartrowcolors{}
 \end{table}

Using profiling, we find that the speed of simdzone is  limited in part by the
\texttt{parse\_rrsig\_rdata} function when parsing the \texttt{.se} zone. It is our main function which processes RRSIG
records, testing for the presence of various tokens (type, algorithm, TTL, time, base64 sequences, etc.) and parsing them as they are encountered. We find that~44\% of all records in the \texttt{.se} zone are of type RRSIG.
The \texttt{.com} zone only contains~5\% of records of type RRSIG\@. In contrast, about half of the records in the \texttt{.com} zone are of type NS and the corresponding function (\texttt{parse\_ns\_rdata}) takes a significant fraction of the running time ($\approx 15$\%).
Irrespective of the chosen zone file, we find that the function \texttt{maybe\_take} is a significant burden corresponding to nearly a third of the running time according to our profiling data. This function advances to the next record, while handling open and closing parentheses and taking care of buffer management.

\section{Conclusion}

Our initial hypothesis is that the good results obtained by the simdjson library for
parsing JSON can be generalized to the parsing of zone files. Our results support
our hypothesis: we reach parsing speeds that are close to a gigabyte per second. The improved performance comes in part from a reduction in the number
of instructions.
 However,
we do not expect that our results are optimal. Further engineering efforts might
increase further our performance.

We have adopted a fast base64 decoder. Yet our decoder does not use SIMD instructions. Future work should examine the benefit of SIMD-optimized base64 decoding~\cite{mula2018faster,mula2020base64} for zone files.

There are also other directions to explore. 
ARM-based processors have become more popular on servers: Amazon's Graviton processors
are one example. Yet currently simdzone only provides acceleration for x64 processors.
Future work should consider the advanced instructions available on ARM processors 
such as SVE and NEON\@.
SIMD code written for x64 processors does not directly translate into optimally efficient ARM NEON or SVE code due to architectural differences, although some software libraries such as SIMDe may provide a fast conversion. For example, there is no direct equivalent to the \opcode{vpmovmskb} instruction.

Furthermore, we should expect that gains are possible if we leverage the AVX-512 instructions available on many server-class x64 processors. The AVX-512 instruction sets not only support wider registers (64~bytes), but they also introduce many new powerful instructions. For example, many of its instructions support masking so that we can load, store or operate on only part of a SIMD register.  The new compression/decompression instructions can dynamically prune or insert values (bytes or words) into existing registers.
Similarly,
though simdzone relies on a single core, future work should consider multicore parallelism~\cite{10.14778/3436905.3436926,pavlopoulou2018parallel}.

Our work focused on the efficient in-memory parsing of zone files. 
Since deploying simdzone in NSD, we received
comments from some independent users testifying that loading their zone files was much faster.\footnote{\url{https://lists.nlnetlabs.nl/pipermail/nsd-users/2024-April/003304.html}}
Yet DNS servers have many functions and parsing is just one component. Future work should quantify the benefits of such parsing in the operation of a DNS server. Further, we expect other tools could benefit from adopting simdzone in the future (e.g., \texttt{ldns-signzone}, \texttt{Unbound}). Future work should quantify the benefits of simdzone within these tools.

\bibliography{simdzone}

\end{document}